\documentclass[preprintnumbers,amsmath,amssymb,prd]{revtex4}

\usepackage{graphicx}
\usepackage{dcolumn}
\usepackage{bm}
\usepackage{subfigure}

\newcommand{\mpc}{\, {\rm Mpc}}

\newcommand{\hmpc}{\, h^{-1} \mpc}
\newcommand{\ihmpc}{\, h\, {\rm Mpc}^{-1}}

\newcommand{\kms}{{\rm km~s^{-1}}}

\newcommand{\lyaf}{Ly$\alpha$ forest}

\newcommand{\bsig}{\bar{T}}
\newcommand{\tsig}{{\tilde{T}}}

\newcommand{\kNL}{k_{\rm NL}}

\newcommand{\vp}{\mathbf{p}}
\newcommand{\vs}{\mathbf{s}}
\newcommand{\vx}{\mathbf{x}}
\newcommand{\vk}{\mathbf{k}}
\newcommand{\vnabla}{\mathbf{\nabla}}
\newcommand{\vv}{\mathbf{v}}
\newcommand{\Pdd}{P_{\delta\delta}\left(k\right)}
\newcommand{\Pdt}{P_{\delta\theta}\left(k\right)}
\newcommand{\Ptt}{P_{\theta\theta}\left(k\right)}
\newcommand{\Pds}{P_{\delta\pi}\left(k\right)}
\newcommand{\Pts}{P_{\theta\pi}\left(k\right)}
\newcommand{\Pss}{P_{\pi\pi}\left(k\right)}

\begin{document}

\title{How to generate a significant effective temperature for cold dark 
matter, from first principles}

\author{Patrick McDonald}
\email{pmcdonal@cita.utoronto.ca}
\affiliation{Canadian Institute for Theoretical Astrophysics, University of
Toronto, Toronto, ON M5S 3H8, Canada}

\date{\today}

\begin{abstract}

I show how to reintroduce velocity dispersion into perturbation theory (PT) 
calculations of structure in the Universe, i.e., how to go beyond the 
pressureless fluid approximation, starting from first principles. This 
addresses a possible deficiency in uses of PT to compute clustering on the 
weakly non-linear scales that will be critical for probing dark energy. 
Specifically, I show how to derive a non-negligible value for the (initially 
tiny) velocity dispersion of dark matter particles, $\left<\delta v^2\right>$, 
where $\delta v$ is the deviation of particle velocities from the local bulk 
flow. The calculation is essentially a renormalization of the homogeneous (zero
order) dispersion by fluctuations 1st order in the initial power spectrum. For 
power law power spectra with $n>-3$, the small-scale fluctuations diverge and 
significant dispersion can be generated from an arbitrarily small starting 
value -- the dispersion level is set by an equilibrium between fluctuations 
generating more dispersion and dispersion suppressing fluctuations. For an 
$n=-1.4$ power law normalized to match the present non-linear scale, the 
dispersion would be $\sim 100~\kms$. This $n$ corresponds roughly to the slope 
on the non-linear scale in the real $\Lambda$CDM Universe, but $\Lambda$CDM 
contains much less initial small-scale power -- not enough to bootstrap the 
small starting dispersion up to a significant value within linear theory 
(viewed very broadly, structure formation has actually taken place rather 
suddenly and recently, in spite of the usual ``hierarchical'' description). The
next order PT calculation, which I carry out only at an order of magnitude 
level, should drive the dispersion up into balance with the growing structure, 
accounting for small dispersion effects seen recently in simulations. 
 
\end{abstract}


\maketitle

\section{Introduction}

Observing the large-scale density fluctuations in the Universe is one of the 
best ways we have to approach many fundamental questions about the Universe,
e.g., understanding inflation, dark matter, dark energy, the curvature 
of the Universe, neutrino masses, possible extra dimensions, 
modifications of gravity, etc. \citep[e.g.,][]{2009ApJS..180..330K,
2005PhRvD..71j3515S,
2006JCAP...10..014S,2006PhRvL..97s1303S,2007PhRvD..76h3004S,
2003ApJ...594L..71A,2009arXiv0907.2257D,2009arXiv0908.2285L,
2009MNRAS.398..321G,2009arXiv0909.0751B,2008PhRvD..78l3534T,
2008arXiv0810.0323M,2008PhRvD..78l3519M,
2009arXiv0906.4545S,
2009arXiv0906.4548C,2009arXiv0907.5220M}.
Statistics of the current density fluctuations can be used to infer
statistics of the small initial perturbations from which they grew, and in turn
to understand the physics of the very early Universe.
Measuring the evolution of large-scale structure (LSS) over time tells us
about the present matter content of the Universe and the dynamical rules its
evolution follows. Before we can learn anything about fundamental properties of
the Universe, however, we must be able to compute the directly observable
astrophysical quantities (e.g., the CMB \cite{2009ApJS..180..296N}, 
galaxy clustering 
statistics \citep{2006PhRvD..74l3507T}, \lyaf\ absorption
\citep{2005ApJ...635..761M,2006ApJS..163...80M,2006MNRAS.365..231V,
2007PhRvD..76f3009M,2003ApJ...585...34M,2002ApJ...580...42M},
galaxy ellipticity correlations used to measure weak lensing 
\citep{2006ApJ...647..116H}, galaxy cluster/Sunyaev-Zel'dovich effect (SZ) 
measurements \citep{2008RPPh...71f6902A}, and possibly
future 21cm surveys \citep{2008PhRvL.100i1303C}) given a 
hypothetical underlying model. 

As observational statistics become more and more
precise, with the potential to measure more and more subtle differences between
models, the requirements on the phenomenological theory needed for their 
interpretation become correspondingly 
more stringent.  Currently, linear-order perturbation theory 
\citep{1980lssu.book.....P,
1996ApJ...469..437S,1984ApJ...284L...9K,1987MNRAS.227....1K}
provides our primary means of 
calculating LSS observables for cosmological parameter estimation and model 
testing, with only 
ad hoc, and recently demonstrably inadequate, attempts to correct for 
non-linearity once it becomes non-negligible 
\citep{2008MNRAS.385..830S,2006PhRvD..74l3507T,2006JCAP...10..014S,
2008JCAP...07..017H,2009ApJS..180..330K} (a vast number of papers have been 
written on beyond-linear calculations, but most of these are never used in
parameter estimation papers).  Linear theory can robustly describe 
observations on very large scales or at very early times, but breaks down when
the perturbations become too large on a given scale.
When considering gravitational evolution only (i.e., dark matter only), 
numerical simulations can be used to compute the fully non-linear evolution 
of the density field to high accuracy (with a lot of care and computer 
power \cite{2008CS&D....1a5003H,2008arXiv0812.1052H,2009arXiv0902.0429H}), 
however, as 
discussed extensively in \cite{2009JCAP...08..020M}, numerical 
simulations are
a less than ideal tool for interpreting future precision observations,
once one considers real observables which are inevitably influenced by baryonic
effects (e.g., star formation, and the accompanying complication of general 
gas dynamics). 
To summarize the argument in \cite{2009JCAP...08..020M}:
Beyond linear order perturbation theory (PT)  
\citep{1980lssu.book.....P,1981MNRAS.197..931J,1983MNRAS.203..345V,
1984ApJ...279..499F,1986ApJ...311....6G,1996ApJ...473..620S,
2002PhR...367....1B,2006ApJ...651..619J} should provide the primary means of
interpreting very high precision future LSS data, just like linear theory has
provided the primary means in the past. The range of scales over which higher 
order PT will 
be necessary (i.e., linear theory is insufficient) and sufficient (i.e., it 
will be accurate enough after computing a modest number of terms) will 
become larger as the precision of observations
increases, while the chances of robustly, completely, convincingly describing 
the full precision of the observations using inevitably somewhat ad 
hoc prescriptions for star formation and related things in simulations becomes 
more remote. The key difference between PT and simulations is the fact that
perturbative clustering can be completely described by a finite set of 
well-defined parameters, no matter how complicated the small-scale 
physics is (at least as conjectured in \cite{2009JCAP...08..020M}),
while the need for fully non-linear calculations implies that there is
generally no bound on the number or form of free parameters (more or less by 
definition).  The idea that the importance of PT relative to simulations is 
increasing with time may seem backwards relative to 
conventional wisdom, however, my argument is that this conventional wisdom was
developed for the era of not very precise observations, when corrections to 
linear
theory were already too large to be described by PT by the time they were 
large enough to be statistically significant, i.e., past PT work was 
premature.  To be clear, I am not saying that
simulations will not be extremely useful, only that they are unlikely to be
the leading tool for extracting fundamental cosmological information from very
high precision observations (much like the situation in high energy physics,
where lattice QCD simulations provide much qualitative, 
and recently even quite precise quantitative, insight 
\cite{2008Sci...322.1224D}, but
high precision constraints on models are made primarily in the regime 
accessible to perturbation theory \cite{2006JPhG...33....1Y}).  

Following the above line of reasoning, this paper is aimed at building up our 
ability to do calculations based on perturbation theory. It deals 
exclusively with the gravity-only part of the calculation, however, this should
be viewed only as an intermediate goal.
PT is most necessary for practical computations of 
observables which can not realistically be done from first principles in a 
simulation --
understanding how to do computations for dark matter-only is simply a 
prerequisite for computing these observables. This continues a recent line of 
work related to 
renormalization/resummation methods \cite{2006PhRvD..73f3520C,
2006PhRvD..73f3519C,2007JCAP...06..026M,2007PhRvD..76h3517I,
2008ApJ...674..617T,2007PhRvD..75d3514M,2008A&A...484...79V,
2008PhRvD..77f3530M,2008MPLA...23...25M,2008PhRvD..77b3533C,
2008JCAP...10..036P,
2008PhRvD..78j3521B,2008PhRvD..78h3503B,
2009JCAP...06..017L,2009MNRAS.397.1275W,2009PhRvD..80d3531C,
2009PASJ...61..321N}. 
In my opinion, the key to maximizing the usefulness of all of
this work is eventually connecting it to galaxy biasing and related 
complications \cite{2005PhRvD..71d3511S,2006PhRvD..74j3512M,
2008PhRvD..78h3519M,2009JCAP...08..020M}.

One traditional potential limitation of LSS perturbation theory, which I focus 
on addressing in this paper, is that it assumes the 
particles at a point in space all have exactly the same velocity
\cite{2009PhRvD..80d3504P}. The equations
used are those of a pressureless fluid. 
This is often referred to as the ``single-stream approximation'', however, I 
will 
avoid this language as it assumes a certain picture of large-scale structure
formation that may or may not have anything to do with reality. I say this  
because, for typically observed times, stream crossing is actually 
ubiquitous -- in fact, there are many orders of magnitude in scale of deeply 
non-linear structure, to the point where the initial conditions on very small
scales may be effectively forgotten. Of course, the standard language 
implicitly 
means no stream crossing when the field is in some sense smoothed on the 
typical scale where the perturbation theory is
supposed to apply. When you look at it this way, however, it is clear that, if 
coldness is a good effective theory, it is not simply because
the particles were initially cold, there is an additional requirement that 
the velocity field remains effectively cold on scales just smaller than the 
ones of interest. 

To give a qualitative preview of the paper:
The exact equation for the evolution of collisionless particles is the 
Vlasov equation for the full distribution function in 6-dimensional 
phase space (particle density in position and momentum space). The
standard hydrodynamic 
equations are derived by taking moments of the 
Vlasov equation with respect to momentum -- the zeroth moment gives 
an equation for the evolution of density, the first moment gives an equation 
for the evolution of bulk velocity, and higher moments are normally 
dropped, e.g., the second moment
which would describe velocity dispersion. At first glance, one 
might think that standard PT could be extended by simply adding the evolution 
equation for the 2nd and possibly higher moments, however, that turns
out to be less straightforward than it sounds.
Viewed conventionally, dropping these higher moments is not really an added 
approximation
in standard Eulerian perturbation theory. The
lowest order evolution equations contain
only a Hubble drag term, so any small initial velocity dispersion will 
rapidly become even smaller.
Higher order equations contain no source terms, meaning that the higher order
results are always proportional to the tiny starting value.
\cite{2001A&A...379....8V} showed that even perturbation theory using 
the full distribution function directly leads to the same result. I will show, 
however, that this argument for dropping dispersion from the 
perturbation theory is faulty, in that, while the lowest order terms are
very small, the series is very rapidly diverging, in the sense that higher 
order terms are increasing in size instead of decreasing. This implies that 
some rearrangement of the series is needed, as in a resummation or 
renormalization group calculation. In case this mathematical reasoning is not 
sufficient motivation,
recently \cite{2009PhRvD..80d3504P} computed the velocity dispersion 
generated in N-body simulations, finding it to be small but not completely 
dynamically negligible. 

There has been much discussion in the past about different ways of adding 
velocity dispersion, or more general changes to the small-scale effective 
theory used for PT calculations 
\cite[e.g.,][]{2004PhRvD..70f4010T,2005PhRvD..71f7302R,
2006PhRvD..73b4024S,2002MNRAS.330..907M,2005A&A...438..443B,
2007JPhA...40.6849G,2009ApJ...700..705S}, 
however, these papers all lacked a first
principles derivation, starting from the exact equations, of the model they 
use. This made their
usefulness for precision calculations questionable, as there were always 
added assumptions and/or free parameters. The point of this paper is to show 
how to do a straightforward computation that takes us directly from the initial
homogeneous, cold, starting theory to a theory with
highly developed, potentially hot, small-scale structure. 

The rest of the paper is as follows:
In \S\ref{seccalc} I show how to use renormalization group-inspired ideas to 
reintroduce velocity dispersion from first principles. 
In \S\ref{secredshiftspace} I give a short discussion of the 
implications of these velocity dispersion calculations for redshift-space 
distortions.
Finally, in \S\ref{secdiscuss} I discuss the results.

\section{Generation of velocity dispersion \label{seccalc}}

In this section I present the calculation that generates velocity dispersion.  
In \S \ref{secevolutioneqs} I discuss the time evolution equations for cold 
dark matter, which are the starting point for perturbation theory.
In \S \ref{secbarepert} I compute the velocity dispersion taking the standard 
PT approach, which leads to negligible dispersion.  
In \S \ref{secRG} I apply an approximate renormalization group approach to 
models with power law initial power spectra, which 
demonstrates how significant dispersion can be generated.
Finally, in \S \ref{secfullcalc} I take a related but more exact approach,
which should be easily extendible to more general, higher order, calculations, 
including $\Lambda$CDM power spectra.

\subsection{Evolution equations \label{secevolutioneqs}}

The exact evolution of collisionless particles
is is described by the Vlasov equation \citep{1980lssu.book.....P}
\begin{equation}
\frac{\partial f}{\partial \tau} +\frac{1}{a~m}~\vp\cdot\vnabla f
-a~m~\vnabla \phi \cdot \vnabla_\vp f=0 ~,
\label{vlasov}
\end{equation}
with 
\begin{equation}
\nabla^2\phi = 4~\pi~G~a^2 \bar{\rho}~\delta=
\frac{3}{2}H_0^2~\Omega_{m,0}~\frac{\delta}{a}~,
\label{potential}
\end{equation}
where $f(\vx,\vp,\tau)$ is the particle density at phase-space position 
$(\vx,\vp)$,
$m$ is the particle mass (which plays no role in the final results), and 
$\vp= a~m\vv_p$ ($\vv_p$ here is a particle's peculiar velocity, not to be
confused with the mean peculiar velocity used everywhere else). 
$\vx$ is the comoving position and 
$\tau=\int dt/a$ is the conformal time, with $a=1/(1+z)$ the expansion factor.
Except when otherwise indicated $\vnabla=\mathbf{\partial/\partial x}$.
The density field is obtained by averaging the distribution function over 
momentum:
\begin{equation}
\rho(\vx,\tau)\equiv m a^{-3}\int d^3 p~f(\vx,\vp,\tau)~,
\end{equation}    
and the bulk (mean) velocity and higher moments of the velocity distribution
e.g., the dispersion of particle velocities around their bulk velocity, can be 
similarly obtained by multiplying the distribution function by any number of
$\vp$'s (e.g., one to obtain bulk velocity) before integrating over $\vp$.
The mean velocity of the particles at $\vx$ is
\begin{equation}
\vv\left(\vx,\tau\right) \equiv \frac{\int \left(\vp/m a\right) f~ d^3p}
{\int f~ d^3p}~.
\end{equation}
and the velocity dispersion tensor is
\begin{equation}
\sigma^{ij}\left(\vx,\tau\right) \equiv
\frac{\int \left(p_i p_j/m^2 a^2\right) f~ d^3p}
{\int f~ d^3p} - v^i v^j~,
\end{equation}
i.e., 
$\sigma^{i j}(\vx,\tau)\equiv\left<\delta v^i~\delta v^j\right>_p$ 
with $\delta v^i$ the deviation of a particle's velocity from the local 
mean velocity, and the average is over all particles at $\vx$.  

As discussed by \cite{1980lssu.book.....P}, taking moments of 
the Vlasov equation with respect to momentum leads to a hierarchy of 
evolution equations for these quantities.
The density evolution equation is the usual continuity equation,
\begin{equation}
\frac{\partial \delta}{\partial \tau}+
\partial_i\left[\left(1+\delta\right)v^i\right]=0~,
\label{eqcontinuity}
\end{equation}
where $\delta(\vx,\tau)=\rho(\vx,\tau)/\bar{\rho}-1$,
$\partial_k=\partial/\partial x^k$.
The bulk velocity evolution equation acquires a velocity dispersion term that 
is usually dropped to give the Euler equation for standard perturbation theory
\begin{equation}
\frac{\partial v^i}{\partial \tau}+\mathcal{H} 
v^i+v^j \partial_j v^i=
-\partial_i \phi 
-\frac{\partial_j\left[\left(1+\delta\right)\sigma^{i j}\right]}{1+\delta}~,
\label{eqeuler}
\end{equation}
where $\mathcal{H}=d\ln a/d\tau=a H$ with $H$ the usual Hubble parameter.
Multiplying the Vlasov equation by $p_i p_j$ and integrating over 
$\vp$ gives (after some substitution to remove bulk velocity terms) 
\begin{equation}
\frac{\partial \sigma^{i j}}{\partial \tau}+2~ {\mathcal H}~\sigma^{i j}+
v^k~ \partial_k \sigma^{i j}+\sigma^{j k}~ \partial_k v^i+
\sigma^{i k}~ \partial_k v^j=
-\frac{\partial_k\left[\left(1+\delta\right) q^{i j k}\right]}{1+\delta} ~,
\label{eqdispersion}
\end{equation}
where 
$q^{i j k}(\vx,\tau)\equiv\left<\delta v^i~\delta v^j~\delta v^k\right>_p$.
Similar equations can be derived for $q^{ijk}$ and higher moments.

\subsection{Standard perturbation theory \label{secbarepert}}

Perturbation theory consists of writing the fields as
a series of terms of at least formally increasing order of smallness, i.e., 
$\delta=\delta_1+\delta_2+\delta_3+...$.  The evolution equations are solved
order-by-order, with lower order solutions appearing as sources in the 
higher order equations so that $\delta_n$ is of order $\delta_1^n$
\citep{2002PhR...367....1B}.

For simplicity, I will often assume an Einstein-de Sitter (EdS) Universe.  
This assumption makes analytic calculations easier without
qualitatively changing the results.
We then have useful relations:
\begin{equation}
a=\left(\frac{\tau}{3 t_0}\right)^2=\left(\frac{\tau H_0}{2}\right)^2,
~~
{\mathcal H}=\frac{2}{\tau},~~{\rm and}~~
\frac{3}{2}\frac{H_0^2}{a}=\frac{6}{\tau^2}~.
\end{equation}

Symmetry allows for a zero order (assumed to be homogeneous and isotropic) 
component of 
$\sigma^{ij}$, 
$\bsig_0(\tau)\equiv
\bar{\sigma}_0^{11}=\bar{\sigma}_0^{22}=\bar{\sigma}_0^{33}$, 
and in fact we expect there to be primordial 
velocity dispersion for realistic WIMPs, albeit very small.
The background dispersion evolves following:
\begin{equation}
\bsig_0(\tau)\propto a^{-2} \propto \tau^{-4}~,
\end{equation}
where the normalization must be fixed by the initial conditions.

The linearized equations are:
\begin{eqnarray}
\frac{\partial \delta_1}{\partial \tau}+
\partial_iv_1^i &=& 0 \\
\frac{\partial v_1^i}{\partial \tau}+\mathcal{H} 
v_1^i&=&
-\partial_i \phi_1 
-\partial_j\sigma_1^{ij}-\bsig_0\partial_i \delta_1 \\
\frac{\partial \sigma_1^{i j}}{\partial \tau}+2~ {\mathcal H}~\sigma_1^{i j}+
\bsig_0~ \partial_j v_1^i+
\bsig_0~ \partial_i v_1^j&=& 0 ~,
\end{eqnarray}
where the Poisson equation holds order by order relating $\delta$ to $\phi$.
Here is where the story ends for velocity dispersion in standard PT.  The CDM 
velocity dispersion is
supposed to be initially small, and only gets smaller.  At 1st order the 
evolution equation for $\sigma^{ij}$ contains only the Hubble drag term and
terms proportional to the tiny $\bsig_0$, so the perturbations will remain 
small.  The last term from
Eq. (\ref{eqeuler}) can be dropped and the hierarchy is closed.  I will show 
that
this treatment is justified in the approach of standard PT, but retaining the
apparently small dispersion terms allows for the renormalization in the next
subsection.  I will drop $q^{ijk}$ for simplicity (and because symmetry 
prevents it from acquiring a non-zero background value).

In this paper my only goal is to renormalize the zero-order velocity 
dispersion, $\bsig_0$, so I am going to henceforth
assume the velocity field is curl-free and only solve for 
$\vnabla \cdot \vv$ and $\partial_i \partial_j \sigma^{i j}$.  The techniques
of this paper could probably be used to reactivate the vorticity variables, but 
symmetry guarantees that they will not have any homogeneous (zero order, i.e., 
background) component (see \cite{2009PhRvD..80d3504P} for a measurement of the
vorticity power spectrum in simulations, and an argument that vorticity is 
completely irrelevant for large-scale clustering).
The linearized equations, assuming EdS, can be re-written as:
\begin{equation}
\label{eqdelta1}
\frac{\partial \delta_1}{\partial \eta}=
\theta_1
\end{equation} 
\begin{equation}
\label{eqtheta1}
\frac{\partial \theta_1}{\partial \eta}+\frac{1}{2}~
\theta_1
=\frac{3}{2}~\delta_1
-2~ k^2 \tsig_0\left(\delta_1+2 \pi_1\right)
\end{equation}
\begin{equation} 
\label{eqpi1}
\frac{\partial \pi_1}{\partial \eta}+\pi_1 = \theta_1
-\frac{\partial \ln \tsig_0}{\partial \eta}~ \pi_1 ~,
\end{equation}
where
\begin{eqnarray}
\eta&=&\ln a\\
\theta &=& \frac{i \vk \cdot \vv}{\mathcal{H}} \\
\pi &=& \frac{1}{2}\frac{k_i k_j}{k^2}\frac{ \sigma^{ij}}{\bsig_0} \\
\tsig_0&=&\frac{\bsig_0}{\mathcal{H}^2}
\end{eqnarray}
I have moved to Fourier space, e.g., 
$\delta(\vk,\tau)=\int d^3\vx~\exp(i \vk\cdot \vx)~\delta(\vx,\tau)$,
where $\nabla\rightarrow -i \vk$. 

These equations can't be solved exactly.
We expect that the density and velocity modes will be as usual 
linear theory at low $k$ until suppressed
below some Jeans-like filtering scale.
The basic behavior can be found more quantitatively from a small $k$ 
expansion, where I first drop the 
$k^2$ terms and obtain, using $\tsig_0=\tsig_i a_i/a$, the usual growing mode 
solutions,
$\delta_1(k\rightarrow 0)=\theta_1(k\rightarrow 0)=\pi_1(k\rightarrow 0)=
\delta_i a/a_i$.
I then substitute these solutions into the $k^2$ terms to obtain, to order 
$k^2$, 
\begin{equation}
\delta_1=\theta_1=\pi_1 = \delta_i\frac{a}{a_i} \left(1-\frac{6}{5}k^2 \tsig_i
\right)~.
\end{equation} 
Here I have eliminated the new integration constants at order $k^2$ by 
requiring that the $\mathcal{O}(k^2)$ corrections are zero at some very 
early initial time $a_i$.
With the assumption that the fields are curl free, we can invert the
$\pi_1$ solution to obtain
\begin{equation}
\sigma^{ij}_1=2\bsig_0 \frac{k_i k_j}{k^2} \delta_1
\end{equation}

If I assume the Jeans-like suppression kernel is Gaussian, i.e., 
$\exp\left(-k^2 R_F^2/2\right)$, we have 
$1-6 k^2 \tsig_i/5=1-k^2 R_F^2/2$ so 
$R_F= (12 \tsig_i/5)^{1/2}$.
Note that I have in this derivation only retained the fasted growing parts of
$\delta_1$, $\theta_1$, and $\pi_1$, which is consistent in standard 
perturbation theory where the initial time can be taken to be arbitrarily 
small, but, as we will see, is dangerous when we start renormalizing. 
Finally, note that as long as $k^2 \tsig_i$ is very small, which it 
will be for
CDM on observable scales (basically by definition), we have changed nothing
by retaining the velocity dispersion terms,
i.e., the standard perturbation theory with zero velocity dispersion is 
self-consistent at this order. 

The 2nd order equations are:
\begin{eqnarray}
\frac{\partial \delta_2}{\partial \tau}+
\partial_i v_2^i + \delta_1 \partial_i v_1^i + v_1^i \partial_i \delta_1&=&0 \\
\frac{\partial v_2^i}{\partial \tau}+\mathcal{H} 
v_2^i+v_1^j \partial_j v_1^i=
-\partial_i \phi_2 
-\partial_j\sigma_2^{ij}&+&\delta_1 \bsig_0 \partial_i \delta_1-
\sigma_1^{ij}\partial_j \delta_1-\bsig_0\partial_i \delta_2  \\
\frac{\partial \sigma_2^{i j}}{\partial \tau}+2~ {\mathcal H}~\sigma_2^{i j}+
\bsig_0~ \partial_j v_2^i+ \bsig_0~ \partial_i v_2^j&=&
-v_1^k~ \partial_k \sigma_1^{i j}-\sigma_1^{j k}~ \partial_k v_1^i-
\sigma_1^{i k}~ \partial_k v_1^j ~.
\end{eqnarray}
For now I am only interested in  
$\bsig_2\equiv \left<\sigma^{11}_2\right>=\left<\sigma^{22}_2\right>=
\left<\sigma^{33}_2\right>=\left<\sigma_2^{ii}\right>/3$, 
which will renormalize $\bsig_0$. 
\begin{equation}
\frac{\partial \bsig_2}{\partial \tau}+2~ {\mathcal H}~\bsig_2+
\frac{1}{3}\left<v^k_1~ \partial_k \sigma^{ii}_1+
2 \sigma^{ik}_1~ \partial_k v^i_1\right>
=0 ~.
\end{equation}
Evaluating this in Fourier space, using the same variable redefinitions as 
above, including $\tsig_2=\bsig_2 \mathcal{H}^{-2}$, gives
\begin{equation}
\frac{\partial \tsig_2}{\partial \eta}+\tsig_2=
\frac{2}{3}\tsig_0 \left<\pi_1\theta_1\right> 
\simeq \frac{2}{3}\tsig_0 \left<\delta_1^2\right> ~,
\label{eqsigma2}
\end{equation}
where deriving the term involving 
$\left<\pi_1\theta_1\right>$ only requires
that $\theta_1$ and $\pi_1$ completely describe the velocities, while the term
involving $\left<\delta_1^2\right>$ requires the above approximate relation 
between $\delta_1$, $\theta_1$, and $\pi_1$. 


Using the right-most side of Eq. (\ref{eqsigma2}), and assuming standard 
linear theory for $\delta$, the solution is:
\begin{equation}
\tsig_2=\frac{1}{3}\tsig_0 \left<\delta_1^2\right>+\frac{c}{a}
\end{equation}
where $c$ is a constant.
If I choose $c$ to make $\tsig_2(a=a_i)=0$, and use the
zero order solution $\tsig_0=A/a$ (remembering that 
$\tsig=\bsig/\mathcal{H}^2$), I finally obtain 
\begin{equation}
\tsig(a)=\frac{A}{a} +\frac{1}{3}
\left[\frac{A}{a} \left<\delta_1^2\right>-
\frac{A}{a_i} \left<\delta_1^2\right>_i
\frac{a_i}{a}\right].
\label{eqbaresigma}
\end{equation}
The standard PT approach would be to drop the term on the right in the
brackets, assuming $a>>a_i$.
From the point of view of standard PT, we could conclude that the result for 
$\bsig_2$ is small, for $\Lambda$CDM, so dropping velocity 
dispersion is justified (while $\left<\delta^2_1\right>$ is large for 
$\Lambda$CDM, it is not 
large enough to overcome the small dispersion and 
make $\left<\delta^2_1\right>\tsig$ large);
however, we have a clear breakdown in the premise of the perturbation theory,
because $\tsig_2/\tsig_0\sim \left<\delta^2_1\right> >>1$.
 
\subsection{Renormalization group approach \label{secRG}}

The problem with the standard calculation outlined in \S\ref{secbarepert}, 
which leads me to renormalization,
is that the 2nd term in Eq.~(\ref{eqbaresigma}) ($\tsig_2$) diverges 
relative to the 
first ($\tsig_0$), increasingly as time progresses.  
One might argue that the correction
is still small, in an absolute sense, but this is nonsense because
there is no reason not to expect the higher order terms in the series to be
even larger, i.e., the truth could be arbitrarily large. 
Fortunately, we have tools to deal with this kind of breakdown in perturbative
solutions of differential equations
\cite{1994PhRvL..73.1311C,1995PThPh..94..503K,1997PThPh..97..179K,
2001PhR...352..219S,2006JPhA...39.8061K}. 

A key observation about perturbation theory is 
that there is inevitable ambiguity in the solution that comes from solving 
a set of differential equations at each order.  We obtain a new set of 
integration
constants at each order, while the initial conditions only determine one set.
This is apparent in Eq. (\ref{eqbaresigma}), where I fixed the 2nd order 
integration constant by the {\it arbitrary} requirement that 
$\tsig_2(a=a_i)=0$.  I can instead fix it to set 
$\tsig_2(a_\star)=0$, where $a_\star$ is some arbitrary time, 
so the solution is
\begin{equation}
\tsig(a)=\frac{A_\star}{a}+\frac{1}{3}
\frac{A_\star}{a} \left[\left<\delta_1^2\right>-
\left<\delta_1^2\left(a_\star\right)\right>
\right] 
\end{equation}
where $A_\star$ must depend on $a_\star$, in order to match the initial
conditions.
The final result should not depend on the arbitrary $a_\star$, so the RG 
equation is obtained by taking the derivative of $\tsig(a)$
with respect to $a_\star$, evaluated at $a=a_\star$, and setting
it equal to zero:
\begin{equation}
\left. \frac{d \tsig}{d a_\star}\right|_{a_\star=a}=0=
\frac{d A_\star}{d a}a^{-1}-\frac{1}{3}A_\star a^{-1}
\frac{\partial \left<\delta_1^2\right>}{\partial a}
\end{equation}
or
\begin{equation}
\frac{d A_\star}{d a}=\frac{1}{3}A_\star 
\frac{\partial \left<\delta_1^2\right>}{\partial a}~.
\label{RGeq}
\end{equation}
Now, $\left<\delta_1^2\right>$ generally depends on $A_\star$ through the 
filtering scale 
$R_F$, as discussed above, but supposed for the moment that the power spectrum 
was not truncated so $\left<\delta_1^2\right>$ is independent
of $A_\star$. The solution to the RG equation would be, long after the initial
time,
\begin{equation}
\frac{A_\star}{A_i}=\exp\left(\frac{\left<\delta_1^2\right>}{3}\right)
\end{equation}
where $A_i$ is the initial $A$.  For a power law with $n>-3$,
$\left<\delta_1^2\right>$ is infinite, so we have infinite growth of the 
velocity dispersion,
but even in the case of $\Lambda$CDM with asymptotic high-$k$ slope slightly
less than -3 (because the primordial slope, from inflation, appears to be less 
than one \cite{2006JCAP...10..014S}), 
$\exp\left(\left<\delta_1^2\right>/3\right)$ will
become enormous.  Furthermore, as discussed in \cite{2007PhRvD..75d3514M}, 
higher order corrections to
the power spectrum generically push the slope to larger values, 
so the result, without filtering, will be even larger when calculated to higher
order, to the point of being practically infinite.  Infinite
velocity dispersion is of course too much --
the point here is simply to demonstrate how the ridiculously small initial 
seed velocity dispersion in CDM can grow into substantial
velocity dispersion later.  

Including the Jeans filtering resulting from the velocity dispersion itself
will regulate the growth of the dispersion.  Recall that the approximate 
filtering scale is 
$R_F=(12 \tsig_i/5)^{1/2}= (12 A/5 a_i)^{1/2}$.  
To be precise, $R_F$ quantifies the frozen-in 
power suppression due to some velocity dispersion present at initial time 
$a_i$, long after this dispersion has redshifted away.  When we in effect 
add new dispersion at time $a_\star$, through the RG equation, the smoothing
will not be instantaneous; however, it takes place quite quickly, so I will
assume I can use 
$R_F= \left[12 A_\star/5 \alpha a_\star\right]^{1/2}$ 
in the RG equation, where $\alpha \sim 3$ is a fudge factor to account for the
lag between the introduction of dispersion and the smoothing of the power
spectrum.
As with the original definition for $R_F$, where I assumed that the smoothing 
is Gaussian,
this approximation means the results should only be trusted at the 
order-of-magnitude level.  I will do a more exact calculation below.
Assuming Gaussian filtering, and power spectrum
$\Delta^2(k,a)=\Delta_p^2(a)(k/k_p)^{3+n}$ (with $n>-3$), the variance is 
\begin{equation}
\left<\delta_1^2\right>= \frac{\Delta^2_p(a)}{2}
\frac{\Gamma\left[\left(3+n\right)/2\right]}{
\left(k_p R_F\right)^{3+n}}~.
\label{powerintegral}
\end{equation}
Using Eq. (\ref{powerintegral}) in Eq. (\ref{RGeq}) I find, well after the
initial time,
\begin{equation}
A=\left[\frac{3+n}{6}\frac{\Delta_p^2\left(a\right)}{2}
\frac{\Gamma\left(\left(3+n\right)/2\right)}
{\left(k_p \sqrt{12/5\alpha a}~\right)^{3+n}}\right]^{\frac{2}{3+n}}. 
\label{eqAresult}
\end{equation}
Recall that $\tsig\left(a\right)=A a^{-1}$. 
This result may not be very illuminating, but it can be re-written in a way 
that makes it very clear:
\begin{equation}
\frac{6}{3+n}=\left<\delta_1^2\left[R_F\left(A\right)\right]\right>~, 
\label{eqfixedvariance}
\end{equation}
i.e., the velocity dispersion simply grows to the point where the Jeans-like
filtering reduces the variance to of order unity, with the exact relation 
dependent on the slope of the power spectrum.

\subsubsection{RG method, 2nd iteration}

The calculation leading to Eq. (\ref{eqfixedvariance}) does not
look entirely self-consistent. 
At various stages in it, I used the
fact that $\tsig_0\propto a^{-1}$, however, in the end, the renormalized 
$\tsig$ is $\propto a^{4/(3+n)}$, growing quickly rather than decaying.
The brute force way to solve this problem would have been to compute RG 
equations 
for the pieces of the calculation that depended on this assumption (e.g., the 
amplitude of $\pi$), and solve them jointly. It is simpler, and will lead to 
self-consistent and enlightening results,
to redo the calculation starting with $\tsig_0 = A a^\alpha$, with 
$\alpha=4/(3+n)$.
The large-scale solutions for $\delta$ and $\theta$ are unchanged, but we now 
have $\pi_1(k\rightarrow 0) = (2+\alpha)^{-1} \delta_1(k\rightarrow 0)$. 
The smoothing estimated from the $k^2$ expansion changes for all three fields:
\begin{equation}
\delta_1 \simeq \left[1-
\frac{2 }{(2+\alpha)}
k^2 \tsig_0  
\frac{(4+\alpha)}{\alpha(5+2\alpha)}
\right]
\delta\left(k\rightarrow 0\right)
\end{equation}
\begin{equation}
\theta_1\simeq \left[1-
\frac{2 (1+\alpha)}{(2+\alpha)}
k^2 \tsig_0  
\frac{(4+\alpha)}{\alpha(5+2\alpha)}
\right]
\delta\left(k\rightarrow 0\right)
\label{eqcarefulthetakernel}
\end{equation} 
\begin{equation}
\pi_1 \simeq \frac{1}{2+\alpha}\left[1-k^2 \tsig_0  
\frac{(4+\alpha)}{\alpha(5+2\alpha)}
\right]
\delta\left(k\rightarrow 0\right)~.
\label{eqcarefulpikernel}
\end{equation} 
Note that there is a qualitative difference in these smoothing kernels relative
to the $\alpha<0$ case in that
$\tsig$ appears instead of $\tsig_i$, i.e., because the comoving Jeans scale is
increasing when $\alpha>0$, the smoothing continues to grow with time rather
than freezing out. 
The numerical factors are also different, with $\theta_1$ being generally 
significantly smoother than $\delta_1$, with $\pi_1$ between them. 
For example, 
for $n=-1.4$,  $R_F^\delta\simeq 0.48~\tsig_0^{1/2}$, 
$R_F^\theta \simeq 0.90~\tsig_0^{1/2}$,
and $R_F^\pi\simeq 0.72~\tsig_0^{1/2}$.
Now, 
\begin{equation}
\left<\pi_1\theta_1\right>= \frac{1}{2+\alpha}\frac{\Delta^2_p(a)}{2}
\frac{\Gamma\left[\left(3+n\right)/2\right]}{
\left(k_p R^{\pi\theta}_F\right)^{3+n}}
\label{eqpithetaintegral}
\end{equation}
where $R^{\pi\theta 2}_F=(R^{\pi 2}_F+R^{\theta 2}_F)/2$ and
$R^\theta_F$ and $R_F^\pi$ are the smoothing scales implied by
Eq. (\ref{eqcarefulthetakernel}) and (\ref{eqcarefulpikernel}). 
The important thing to recognize about 
$\sigma^2_{\pi\theta}=\left<\pi_1\theta_1\right>$ here is
that it is time independent, i.e., the growth of the power spectrum is 
canceled by the change in $R^{\pi\theta 2}_F\propto \tsig_0$.

Now, I have recomputed the linear equations given $\tsig_0\propto a^\alpha$,
so it remains only to recompute the perturbative corrections $\tsig_2$.
Eq. (\ref{eqsigma2}) for $\tsig_2$ is changed, because $\tsig_0$ is no longer 
a solution to 
the zero order equation for $\tsig$. This is a key fact -- at the heart
of all renormalization is the concept that the naive lowest order 
result is
not always the best one to perturb around when doing computations to higher 
order. Sometimes it is better to perturb around something else, with the 
criteria for ``better'' being simply that the corrections remain small. The
usual renormalization procedure of doing the calculation to lowest order, 
using the results to compute higher order
corrections, then absorbing large corrections back into the lowest order, is 
an algorithmic way of accomplishing what one could accomplish by simply
perturbing around the better starting point from the beginning (if one has a
way to determine/guess what it is).
Here, I am perturbing around $\tsig_0=A a^\alpha$, i.e., $\tsig_2$ is 
defined by $\tsig=A a^\alpha+\tsig_2+...$, and the evolution equation I 
derive for $\tsig_2$ is:  
\begin{equation}
\frac{\partial \tsig_2}{\partial \eta}+\tsig_2=
\frac{2}{3}\tsig_0 \left<\pi_1\theta_1\right> 
-\left(\alpha+1\right)\tsig_0 ~.
\label{eqT2it2}
\end{equation}
The 2nd term on the right hand side is new.
Note that I have not yet fixed $A$. Normally, $A$ would be set by the initial 
conditions, but there is no need for that
because the homogeneous part of the solution for 
$\tsig_2$ can be used to match the initial conditions 
(then, since the homogeneous solution is just $\propto a^{-1}$, 
the memory of initial conditions will fade away).
If I choose $A$ to make $\frac{2}{3} \left<\pi_1\theta_1\right>
-\left(\alpha+1\right)=0$, which is possible because $\sigma^2_{\pi\theta}$
depends on $A$ through the smoothing kernels, the equation for $\tsig_2$ 
becomes trivial, just the homogeneous equation, i.e., there are no perturbative
corrections to 
$\tsig_0=A a^\alpha$!
(That is, for the correct amplitude, and the special value $\alpha=4/(3+n)$.) 
This result should not be viewed as some
kind of fortuitous coincidence, or artificial tuning. It is what we aim for 
in a renormalization 
group calculation, and reflects the physical correctness of the idea that 
the velocity dispersion should be rapidly growing, tracking the non-linear 
scale, with the the unstable cold start completely irrelevant (at least after
enough time has passed, and for this kind of power law power spectrum).

We immediately have an equation like Eq. (\ref{eqfixedvariance}), except 
coming from a much more accurate calculation:
\begin{equation}
\left<\pi_1\theta_1\right>=
\frac{3}{2}\left(\alpha+1\right)=\frac{6}{3+n}+\frac{3}{2}~.
\end{equation}
The variance of $\delta_1$ is larger, because of the small coefficient of
$\pi_1$, and lesser smoothing of $\delta_1$, i.e., the more accurate version
of Eq. (\ref{eqfixedvariance}) is:
\begin{equation}
\left<\delta_1^2\left[R^\delta_F\left(A\right)\right]\right>=(2+\alpha)
\left(\frac{R_F^{\pi \theta}}{R^\delta_F}\right)^{3+n}
\left<\pi_1\theta_1\right>
=3\frac{\left(5+n\right)\left(7+n\right)}{\left(3+n\right)^2}
\left(\frac{6+n}{3+n}\right)^{\left(3+n\right)/2}~,
\end{equation}
e.g., this evaluates to $\sigma^2_\delta=55$ at $n=-1.4$, with
$\sigma^2_{\pi \theta}=5.2$.
Keep in mind that this is still essentially the linear theory variance.
There will be non-linear corrections, however, especially for relatively
large $n$, it seems like a very good thing to be starting with finite linear
variance rather than the infinite variance of the bare power law. 
For example, for $n>-1$ the standard 1-loop PT correction diverges
\cite{1996ApJ...473..620S}, while here that problem is clearly solved, i.e., we
have a natural high-$k$ cutoff.
Choosing the pivot point at $\kNL$, defined by $\Delta^2(\kNL)=1$, these 
equations can be solved to give
\begin{equation}
\kNL R_F^\delta = \left(\frac{\Gamma\left[\left(3+n\right)/2\right]}
{2~\sigma^2_\delta}\right)^{\frac{1}{3+n}}
=\left(\frac{\left(3+n\right)^2
\Gamma\left[\left(3+n\right)/2\right]}{6\left(5+n\right)\left(7+n\right)}
\right)^{\frac{1}{3+n}} \left(\frac{3+n}{6+n}\right)^{1/2}
\end{equation}
or
\begin{equation}
\kNL \tsig_0^{1/2} = \left(\frac{\left(3+n\right)^2
\Gamma\left[\left(3+n\right)/2\right]}{6\left(5+n\right)\left(7+n\right)}
\right)^{\frac{1}{3+n}} \left(\frac{\left(5+n\right)\left(23+5 n\right)}
{2\left(3+n\right)\left(4+n\right)\left(6+n\right)}\right)^{1/2}~,
\label{eqkNLsigRG}
\end{equation}
e.g., $\kNL \tsig_0^{1/2}=0.12$ for $n=-1.4$. As one might guess, the 
velocity dispersion scale increases relative to the non-linear scale as the
power law increases to include more small-scale power.
Note that this calculation was self-consistent, without any very ugly 
approximations, although the use of
Gaussian smoothing kernels based on the $k^2$ expansion makes the results still
only approximate. 

\subsection{A more exact, general approach\label{secfullcalc}}

A simpler way approach the split between homogeneous background and 
perturbations, which I could have just started with (except I think
the above derivation has some pedagogical value), is to define 
$\sigma_{i j}\left(\tau,\vx\right)=
\left<\delta v^i~\delta v^j\right>_p\left(\tau,\vx\right) -
\bsig\left(\tau\right)\delta^K_{i j}$, with 
$\bsig\left(\tau\right)\delta^K_{i j}\equiv 
\left<\left<\delta v^i~\delta v^j\right>_p\left(\tau,\vx\right)\right>_\vx$.
Then it is clear that 
\begin{equation}
\label{eqTnonlinear}
\frac{\partial \bsig}{\partial \tau}+
2~ {\mathcal H}~\bsig=-
\frac{1}{3}\left<v^k~ \partial_k \sigma^{i i}+2\sigma^{i k}~ \partial_k v^i
\right> ~,
\end{equation}
and
\begin{equation}
\label{eqsigijnonlinear}
\frac{\partial \sigma^{i j}}{\partial \tau}+2~ {\mathcal H}~\sigma^{i j}+
\bsig\left(\partial_i v^j+\partial_j v^i\right)=
-\Delta\left[v^k~ \partial_k \sigma^{i j}+\sigma^{j k}~ \partial_k v^i+
\sigma^{i k}~ \partial_k v^j\right]
\end{equation}
where $\Delta[f]\equiv f-\left<f\right>$
and I have dropped the $q^{i j k}$ term. These are the fully non-linear
equations, with the need to include the right hand side of 
Eq. (\ref{eqTnonlinear})
as the source of homogeneous velocity dispersion, while subtracting the same
thing from 
the perturbations, being a simple result of the definition of the mean and 
perturbations, rather than some kind of renormalization. Either way, however,
we have the same basic idea that the perturbations feed back on the background,
new relative to standard PT with just density and velocity.

\subsubsection{RG approach at the level of equations rather than solutions}

We still need some perturbative approach to solving the coupled equations for
$\bsig$ and the fluctuations in density, velocity, and velocity dispersion. 
The RG approach used above and in \cite{2007PhRvD..75d3514M} is not ideal.
First,
one needs to solve the perturbative equations analytically, which isn't 
generally possible without approximations that can lead to uncontrolled errors.
Second, the RG equation that is derived may not be 
analytically solvable, as I found in \cite{2007PhRvD..75d3514M}. We can solve
the first problem (needing analytic solutions for the perturbation evolution),
without necessarily making the second problem any worse, by considering 
applying the ideas
behind the RG method at the level of equations rather than solutions. 
Suppose we want to solve 
\begin{equation}
\dot{\delta}=f(\delta)
\end{equation}
The standard perturbative equations are
\begin{equation}
\dot{\delta}_1 = f^\prime \delta_1 ~,~~~~ 
\dot{\delta}_2 = f^\prime \delta_2+\frac{f^{\prime\prime}}{2} \delta_1^2  ~,
~~~~ ...
\end{equation}
etc., where $f^\prime = df/d\delta(\delta=0)$, etc..  
Our solution up to 2nd order is $\delta_1+\delta_2$, but note that we could 
solve a different pair of equations like this:
\begin{equation}
\dot{\delta}^\star_1 = f^\prime \delta_1+\Delta \dot{\delta} ~,~~~~ 
\dot{\delta}^\star_2 = f^\prime \delta_2+\frac{f^{\prime\prime}}{2} \delta_1^2 
-\Delta \dot{\delta} ~,
\end{equation}
and get the same solution $\delta_1+\delta_2=\delta_1^\star+\delta^\star_2$. 
The idea, analogous to the above use of the RG method, is to choose
$\Delta \dot{\delta}$ to be 2nd order and absorb any undesirable parts of 
$\delta_2$. If we choose $\Delta \dot{\delta} =
\frac{f^{\prime\prime}}{2} \delta_1^2+
f^\prime\left(\delta^\star_1-\delta_1\right)$
we have
\begin{equation}
\dot{\delta}^\star_1 = f^\prime \delta^\star_1+
\frac{f^{\prime\prime}}{2} \delta^{\star 2}_1~,~~~~ 
\dot{\delta}^\star_2 = f^\prime \delta^\star_2 ~,
\end{equation}
where I have dropped 3rd order terms. Now, if desired, we can set the initial
value of $\delta^\star_2$ to zero and forget about it. All we have left is the
equation for $\delta_1^\star$. This looks like nothing more than a very 
long-winded
way of saying ``if you want to solve $\dot{\delta}=f(\delta)$ to 2nd order, 
solve the truncated equation $\dot{\delta}=f^\prime \delta+
\frac{f^{\prime\prime}}{2} \delta^2$'', and at the level that I will use it 
here, that is really all it is. However, there was potential for 
more than that within the argument, in that $\Delta\dot{\delta}$ could have 
been chosen to be something else if this was convenient to, say, remove a 
particularly divergent part of $\delta_2$ while still producing an easy to 
solve 
equation for $\delta^\star_1$. Also, $\delta_2^\star$ could in principle have
been given some part of the initial conditions. I present this mainly as an 
explanation of the connection between the RG method and the method of simply
numerically solving truncated equations. The difference between solving these
truncated non-linear equations and the original perturbation equations is that
all of the feedback between 2nd and 1st order is included, rather than 
leaving the first order solution fixed while computing the 2nd order solution.

As suggested in \cite{2007PhRvD..75d3514M}, and implemented in 
\cite{2008JCAP...10..036P}, one approach to computing the 
power spectrum of the fields is to use the time evolution equations for the 
fields
to derive time evolution equations for the power spectrum, and to
solve those numerically. Unfortunately, these equations involve the bispectrum,
which must then be solved for, and the bispectrum evolution equations involve
the trispectrum, etc., i.e., one has an infinite hierarchy of equations. 
The method of truncating this hierarchy by setting the connected trispectrum
to zero is equivalent to applying the above reasoning about truncating the 
series of perturbative equations, i.e., at linear order in the initial power
spectrum the perturbative equations only contain the power spectrum, at 2nd
order they only contain the power spectrum and bispectrum, while the 
trispectrum is needed at 3rd order. This type of approach seems like the
only option for including the velocity dispersion as discussed in this paper in
high precision calculations, because one cannot solve the evolution equations
analytically when including the velocity dispersion. One has to recognize that 
standard 
PT could not even produce the analytic results that it does 
(for time evolution -- there are 
generally numerical integrals over $\vk$) without the special fact that, 
to a very good approximation, one can use the solution to the evolution 
equations in an Einstein-de Sitter Universe in a realistic Universe as well --
this is a very fragile situation and any added complication tends to produce
un-solvable equations (e.g., non-negligible massive neutrinos are a good 
example of this, where, additionally, equations including velocity dispersion 
would naturally appear explicitly 
\cite{2009JCAP...06..017L,2009arXiv0907.2922S}),
and this approximation can never describe the kind of dependence on the 
equation of state of 
dark energy found by \cite{2006MNRAS.366..547M}.  

In anticipation of eventually incorporating the velocity dispersion into a 
scheme for evolving the power spectrum and higher order statistical equations
(the alternative method of closing the hierarchy in \cite{2008ApJ...674..617T} 
may also be a useful way to do this), I write here the evolution equations
at 1st order in the power spectrum. Without the new velocity dispersion parts,
these would just be the equations for the standard linear theory power 
spectrum. (I am still assuming an EdS Universe, although that is not necessary 
here.) 
\begin{equation}
P^\prime_{\delta\delta}\left(k\right)=2 \Pdt
\label{eqPddevolution}
\end{equation}
\begin{equation}
P^\prime_{\delta\theta}\left(k\right)= 
\frac{3}{2} \Pdd-\frac{1}{2}\Pdt+\Ptt-k^2 \tsig\left[\Pdd+2 \Pds\right]
\end{equation}
\begin{equation}
P^\prime_{\theta\theta}\left(k\right)=
3\Pdt-\Ptt-2 k^2 \tsig\left[\Pdt+2 \Pts\right]
\end{equation}
\begin{equation}
P^\prime_{\delta\pi}\left(k\right)=
\Pdt + \Pts - \left(1+\frac{d \ln \tsig}{d\eta}\right)\Pds 
\end{equation}
\begin{equation}
P^\prime_{\theta\pi}\left(k\right)=
\Ptt +\frac{3}{2} \Pds-\left(\frac{3}{2}+\frac{d \ln \tsig}{d\eta}\right)\Pts
-k^2 \tsig \left[\Pds+2 \Pss\right]
\end{equation}
\begin{equation}
\label{eqPppevolution}
P^\prime_{\pi\pi}\left(k\right)=
2 \Pts -2 \left(1+\frac{d \ln \tsig}{d\eta}\right)\Pss 
\end{equation}
\begin{equation}
\tsig^\prime+ \tsig=
\frac{2}{3}~\tsig~\frac{1}{2 \pi^2}\int_0^\infty dq~q^2 
P_{\theta \pi}\left(q\right)
\label{eqtsigevolution}
\end{equation}
Intuitive understanding of this last term is that it measures the tendency 
for velocities to converge on overheated places, and vice versa (remember 
that, for the definitions we're using, positive $\theta$ and $\pi$ mean, 
respectively,
convergence and, in a one-dimensional sense at least, negative 2nd derivative
of the dispersion).
Note that, for this paper, writing all of these equations involves some  
redundancy, in that one could just solve Eqs. (\ref{eqdelta1}-\ref{eqpi1}) to
get growth factors as functions of $k$, using them to evaluate 
Eq. (\ref{eqtsigevolution}). When one goes to higher order, however, where 
$\delta$, $\theta$, and $\pi$ are no longer perfectly correlated, one will need
all the equations.

\subsubsection{Power law power spectra}

Before I solve these equations completely numerically, it is informative to
assume a power law power spectrum, for which we can derive some scalings.
Defining $\kNL$ by $\Delta_{\rm L}^2(\kNL)=1$, and using time evolution  
$\Delta_L(k) \propto D^2$, I find $\kNL \propto D^{-2/\left(3+n\right)}$.
The velocity dispersion must follow 
\begin{equation}
\tsig \propto \kNL^{-2} \propto D^\frac{4}{3+n}~.
\end{equation}
Inevitably, this is the same evolution that the RG calculation above predicted.
For this to be consistent with Eq. (\ref{eqtsigevolution}), we must have
\begin{equation}
\left<\theta \pi \right>=\frac{6}{3+n}+\frac{3}{2}~,
\label{eqthetapipowerlaw}
\end{equation}
i.e., as structure grows on large scales, the Jeans-like effect of velocity
dispersion must erase enough small-scale structure to maintain fixed
$\left<\theta \pi \right>$, with the fixed value being larger for power 
spectra where there is less small-scale power. Again, this result is identical
to the RG result.
This applies only for $n>-3$ of course -- for $n<-3$ this discussion breaks 
down right at the beginning, with the definition of $k_{\rm NL}$.
We can also solve analytically for the growing mode amplitude of $\pi$ in 
the $k\rightarrow 0$ limit, 
which is suppressed relative to $\delta$ and $\theta$ because of 
the $d \ln \tsig /d\eta$ factor (this came from the fact that $\pi$
fluctuations are measured relative to the changing background dispersion). 
The result, again the same as in the RG calculation, is
\begin{equation}
\pi(k\rightarrow 0) = \frac{1}{2}\frac{3+n}{5+n} 
\delta\left(k\rightarrow 0\right) ~,
\label{eqpioverdeltapowerlaw}
\end{equation}
i.e., the dispersion fluctuations are smaller for models with less small-scale
power, although the best way of looking at this is probably to think of these
models as the ones where the non-linear scale is changing more quickly, and
therefor the homogeneous dispersion is increasing more quickly, diluting the
perturbations. 
All that remains to be calculated (for a power law power spectrum) are the 
high-$k$ suppression kernels that
must be applied to each field, which generally must be obtained numerically.

\begin{figure}
\subfigure{\includegraphics[width=0.49\textwidth]{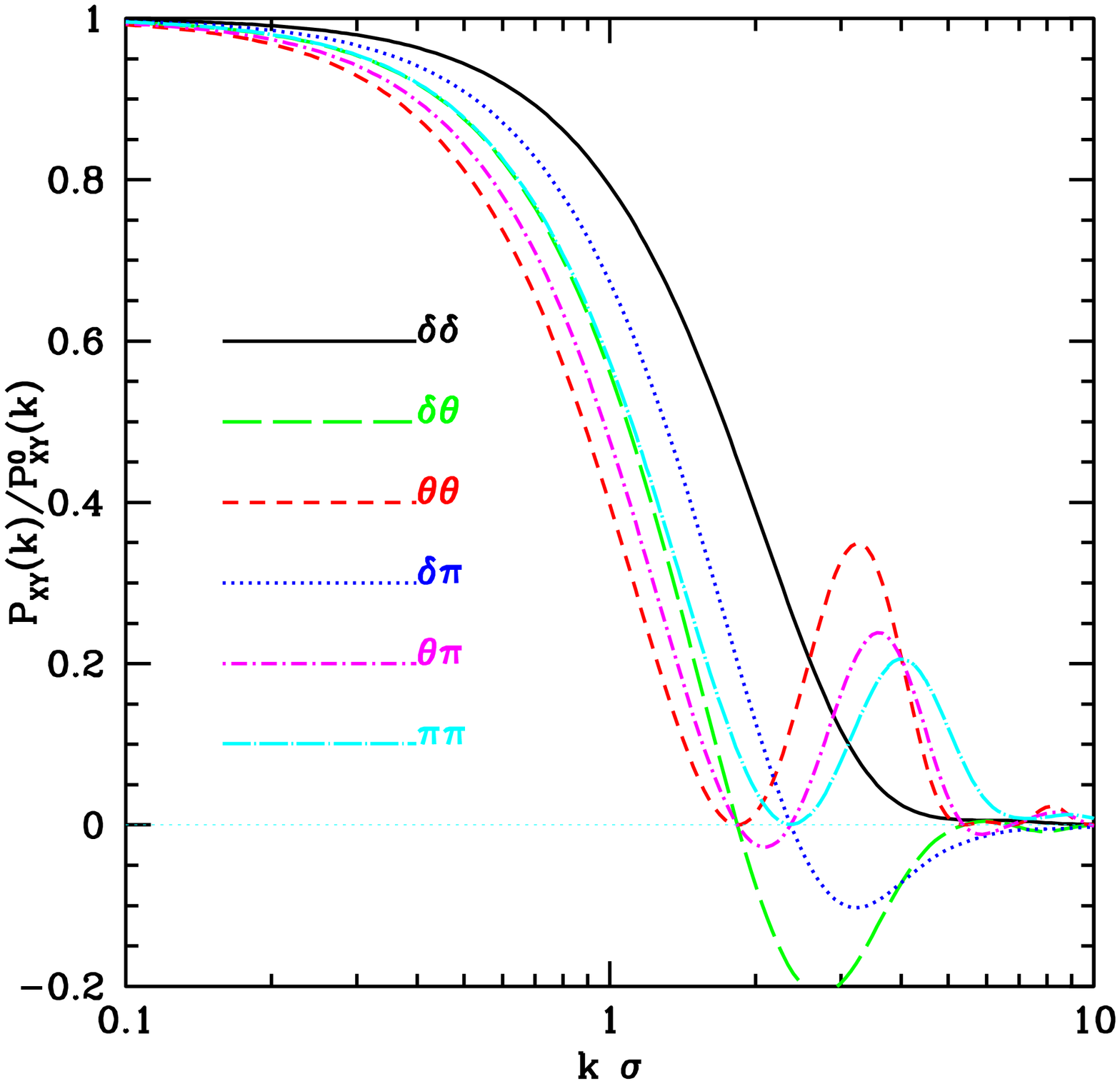}}
\subfigure{\includegraphics[width=0.49\textwidth]{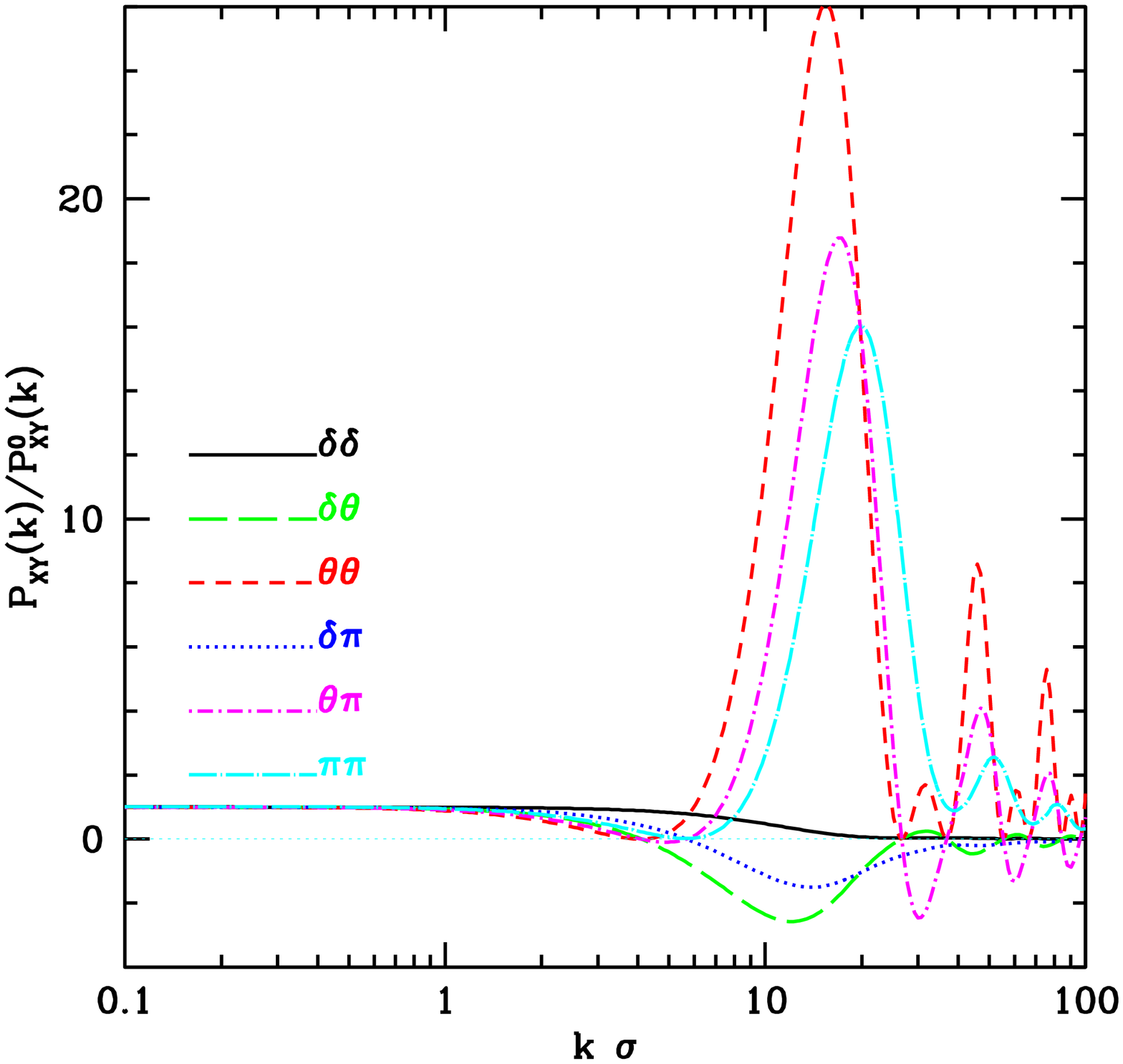}}
\caption{
Ratio of linear power with Jeans-like suppression by velocity dispersion to 
without (i.e., to the $\tsig\rightarrow 0$ limit). 
The left panel is for a power 
law power spectrum with $n=-1.4$, while the right is for $n=-2.75$. 
Lines are identified in the figure panel. $\sigma\equiv \tsig^{1/2}$.
}
\label{figkerns}
\end{figure}

Figure \ref{figkerns} shows these kernels, i.e., the effect of the Jeans-like 
smoothing due to
the mean velocity dispersion, for power law power spectra with $n=-1.4$, 
corresponding to roughly the present non-linear scale, and $-2.75$, 
appropriate to the non-linear scale at a much earlier time. 
To be clear, Fig. \ref{figkerns} was made by evolving Eqs.
(\ref{eqPddevolution}-\ref{eqtsigevolution}) numerically, with initial 
conditions for a scaling solution determined by first evolving assuming the 
scaling solution for $\tsig$, then restarting using the results for the initial
kernels and normalization (at that point, the system will evolve stably on the
scaling solution).   
Generally, the initial conditions would break the
scaling, e.g., if one starts with the bare un-smoothed power law, that is not
consistent with scaling (one would
think that non-linear effects would erase the memory of a break in scaling in 
the distant past, but that does not happen at linear order). 
In each case Fig. \ref{figkerns} shows
the power spectrum from the numerical calculation divided by the power 
spectrum that would be
obtained by taking the $\tsig\rightarrow 0$ limit (specifically, taking the 
overall normalization of $\tsig$ to zero, which does not affect 
$d\ln\tsig/d\eta$ which sets the normalization of $\pi$).
The kernels for the different power laws are dramatically different when 
plotted as functions of 
$k \sigma$ ($\sigma\equiv \tsig^{1/2}$). 
Additionally, the relation between the non-linear scale and 
$\sigma$ is much different: $\sigma= 0.19 k_{\rm NL}^{-1}$ for $n=-1.4$, while
$\sigma=0.000023 k_{\rm NL}^{-1}$ for $n=-2.75$, i.e., the dispersion scale 
lags much farther
behind the non-linear scale for the spectrum with less small-scale power, 
not surprisingly. Note that the non-linear and dispersion scales are increasing
extremely rapidly for $n=-2.75$, like $a^8$, so that the dispersion scale only 
lags the 
non-linear scale by a factor of $\sim 1/4$ in expansion factor. Similarly, even
though the extreme-looking spike in the kernel covers a factor $\sim 2$ in 
$k$, a given mode only spends $\sim 10\%$ of an expansion factor within this 
feature.   
In spite of the apparently large differences, there is a remarkable relation 
between the two power law cases: the expansion factor by which the dispersion 
scale lags the non-linear scale, i.e., 
$\left(\sigma~k_{\rm NL}\right)^{\left(3+n\right)/2}$, is essentially identical
between the two cases -- 0.262 vs. 0.265 from my numerical solution, for 
$n=-1.4$ and $n=-2.75$. To be clear, the dispersion in each case is determined
by the requirement that Eq. (\ref{eqthetapipowerlaw}) for 
$\left<\theta\pi\right>$ is satisfied after the power is filtered by the 
complex-looking kernels in Fig. \ref{figkerns}. 
Eq. (\ref{eqkNLsigRG}) actually does a reasonable job anticipating this 
relation, in spite of its approximations, predicting 0.18 and 0.12
for $n=-1.4$ and $n=-2.75$, respectively, but it
is not clear if there is any deep reason for the near perfect agreement in the
exact calculation (the under-prediction by Eq. \ref{eqkNLsigRG} is 
understandable, as the Gaussian approximation I used there for the kernels 
works essentially perfectly for the initial suppression, but misses the 
wiggles, which produce extra dispersion).

\subsubsection{$\Lambda$CDM}

The situation for a $\Lambda$CDM power spectrum is less straightforward than
for power law power spectra, where we could imagine the initial dispersion was
arbitrarily small, yet still obtain an interesting dispersion tied to the 
non-linear
scale. In the real Universe, we have specific physically motivated initial 
conditions, and we cannot ignore them (at least
not within the lowest order calculation in this paper -- as I discuss below,
I actually expect that the 
influence of the smallest scale initial conditions will be erased when the 
calculation is taken to the next order).
A typical WIMP with mass $m\sim 100$ GeV decouples thermally from the 
relativistic electrons, 
positrons, and neutrinos at $T_d\sim 20$ MeV \cite{2006PhRvD..74f3509B}.
The coupling to the relativistic plasma actually produces acoustic oscillations 
in the transfer function, which I will ignore. 
The 1D velocity dispersion at kinetic decoupling  is $v^{\rm 1D}_d=
\left(\frac{2}{3}\frac{T_d}{m}\right)^{1/2} c$.
For a typical WIMP, the 
scale below which power is erased is of order the scale of free-streaming after 
decoupling, but 
would actually be similar even for much more massive (i.e., slow moving) 
particles, because a substantial fraction of the effect is due to 
Silk-damping-like friction between the WIMPs and leptons during decoupling
\cite{2006PhRvD..74f3509B}. Using Eq. (48) of \cite{2006PhRvD..74f3509B}, I 
will assume the power entering the matter 
dominated era is suppressed by $\exp(-k^2/k^2_{fs})$ with 
$k_{fs}^{-1}\sim \left(\frac{6}{5}\frac{T_d}{m}\right)^{1/2} \tau_d 
\ln\left(\tau_{eq}/1.05 \tau_d\right)$ where $\tau_d$ is the conformal time
at decoupling and $\tau_{eq}$ is the conformal time at matter-radiation 
equality (after which the free streaming has frozen).
For the parameters I am using, the primordial velocity dispersion, 
in Hubble flow distance units, 
extrapolated to the present time assuming no enhancement by structure 
formation, would be $\tsig^{1/2}=8.6\times 10^{-10} \hmpc$, and the 
smoothing scale is $k_{fs}^{-1}=10^{-6} \hmpc$.

The total density variance in a realistic $\Lambda$CDM model, in linear theory,
without any enhancement of velocity dispersion by structure formation, is 
$\left<\delta_1^2\right>\simeq 517 D^2$, where $D$ is the linear theory growth 
factor normalized to be 1 at the present time. If we could assume that 
$\sigma^2_{\theta\pi}=\left<\theta_1 \pi_1\right>$ 
followed the same form, as it does before there is any feedback (and assume an 
EdS Universe),
the solution to Eq. (\ref{eqtsigevolution})
would be $\tsig/\tsig_i=\exp\left[517\left(a^2-a_i^2\right)/3
-\ln\left(a/a_i\right)\right]$. 
We see that if $\left<\theta_1 \pi_1\right>$, which starts out equal to
$\left<\delta_1^2\right>$, remains anywhere near as large, the current 
dispersion
would be extremely large, i.e., $\tsig/\tsig_i\sim \exp(517/3)\sim 10^{75}$.
Clearly this is too large, which is primarily explained by the fact that, once
the velocity dispersion begins to increase, $\left<\theta_1 \pi_1\right>$ falls
well below $\left<\delta_1^2\right>$, as we saw for power laws in 
Eq. (\ref{eqpioverdeltapowerlaw}). We can estimate the size of this effect
by continuing to ignore any additional smoothing by the extra dispersion, but
solving Eqs. (\ref{eqthetapipowerlaw}) and (\ref{eqpioverdeltapowerlaw}) to
find an effective $n$ that will allow both equations to be satisfied, given the
large density variance $\left<\delta_1^2\right>$. The result is 
$\left<\theta_1 \pi_1\right>\simeq 
\left(\frac{3}{2}\left<\delta_1^2\right>\right)^{1/2}$, 
or, for
$\left<\delta_1^2\right>=517 D^2$, $n=-2.77$ (at $D=1$) and 
$\left<\theta_1 \pi_1\right>=27.1 D$. Using this we can solve 
Eq. (\ref{eqtsigevolution}) to obtain  
$\tsig/\tsig_i \sim \exp[(2/3) 27]\sim 10^8$. This may seem like a big number,
but it is in fact inadequate to produce a dispersion that is dynamically 
significant at late times, i.e., we need a boost by $\sim 9$ orders of 
magnitude in $\tsig^{1/2}$, not just 4.
A full numerical evaluation of the evolution equations for $\tsig$ and the
power spectra basically confirms this analysis. $\tsig^{1/2}=6\times 10^{-6}
\hmpc$
at the present time, $\sim 4$ orders of magnitude larger than the primordial
dispersion but nowhere near macroscopic, and not even large enough to begin
suppressing the power much beyond the primordial $k_{fs}^{-1}=10^{-6} \hmpc$.
The full calculation gives
$\left<\delta_1^2\right>=506$ and $\left<\theta_1 \pi_1\right>=26.4$, close to 
the no-smoothing approximation. 
 
Now, the derivation of Eq. (\ref{eqtsigevolution}) did not assume linear 
theory, only that velocities are irrotational, and that we can drop higher 
moments of the velocity distribution function, so it would be correct to 
use a non-linear version of $\left<\theta \pi\right>$ to evaluate it. For 
example, the 
standard 1-loop PT power spectrum has $\left<\delta^2\right> >10^5$, easily 
large enough to completely change the picture for the growth of dispersion, 
i.e., the feedback between the production of dispersion by non-linearity and
the smoothing of the perturbations by dispersion will be critical. 
Exactly how this works out is impossible to say at this point. We don't
have higher order PT calculations that include this form of velocity 
dispersion, and,
as I found above even at linear order, it is almost surely impossible to derive 
analytic PT results with dispersion. The next step appears to be to set up and
solve numerically the
non-linear versions of Eqs. (\ref{eqPddevolution}-\ref{eqPppevolution}), and
the inevitably accompanying bispectrum equations (at least). 
It seems very likely that this will produce the level of dispersion necessary
to produce the percent level corrections to the power spectrum at 
$k \sim 0.1 \ihmpc$ found by \cite{2009PhRvD..80d3504P}.
This requires a dispersion level corresponding to $\sim 1\hmpc$. For 
non-linear scale $k_{\rm NL}$, and $n\sim -1.4$, the result I found for a power
law power spectrum was just this $\sim 1\hmpc$. This would also be consistent
with the scale at which the RG evolution in \cite{2007PhRvD..75d3514M}
saturated to it's fixed-point slope of $n=-1.4$.

\section{Redshift-space distortions \label{secredshiftspace}}

Interpretation of galaxy redshift surveys generally requires us to account for 
the distortion of the observed clustering due to Doppler shifts caused by 
peculiar velocities \cite{1987MNRAS.227....1K,2001MNRAS.327..577B,
2004PhRvD..70h3007S,2005MNRAS.358..939P,
2008PhRvD..78j3512S}, including, 
potentially, those due to velocity dispersion
as well as bulk flows. Here I will continue to discuss dark matter only -- the 
differences between galaxies and dark matter, while certainly important, are 
the subject for another paper.

Starting from the distribution function, defined above, it is straightforward 
to derive the
perturbative redshift-space distortions including possibly velocity dispersion.
Defining the redshift-space coordinate for a particle, 
$\vs=\vx+\hat{r}~p _\parallel/a m \mathcal{H}$
where $\hat{r}$ is the unit vector pointing along the observer's line
of sight and $p_\parallel = \vp\cdot \hat{r}$,
the density in redshift space is:
\begin{equation}
\rho_s\left(\vs\right)=
m~ a^{-3} \int d^3\vx~ d^3\vp~ f\left(\vx,\vp\right)
\delta^D\left(\vs-\vx-\hat{r}\frac{p_\parallel}{a m \mathcal{H}}\right)=
m~ a^{-3} \int d^3\vp~ f\left(\vs-\hat{r}
\frac{p_\parallel}{a m \mathcal{H}},
\vp\right)~.
\end{equation}
I now expand the distribution function in a Taylor series in 
$p _\parallel/a m \mathcal{H}$,
\begin{equation}
\rho_s\left(\vs\right)=
m~ a^{-3} \int d^3\vp f\left(\vs,\vp\right)+
m~ a^{-3} \int d^3\vp \frac{\partial f\left(\vs,
\vp\right)}{\partial s_\parallel}
\left(-\frac{p_\parallel}{a m \mathcal{H}}\right)+
\frac{1}{2} m~ a^{-3} \int d^3\vp \frac{\partial^2 f\left(\vs,
\vp\right)}{\partial s_\parallel^2}
\left(\frac{p_\parallel}{a m \mathcal{H}}\right)^2
+...
\end{equation}
Taking the $s_\parallel$ derivatives outside of the $\vp$ integrals,
and using the above relations between moments of the distribution function
with respect to momentum and bulk velocity and velocity dispersion, I obtain
\begin{equation}
\delta_s\left(\vs\right)=\delta\left(\vs\right)-\frac{1}{\mathcal{H}}
\frac{\partial}{\partial s_\parallel}\left[\left(1+\delta\right)v_\parallel
\right]+\frac{1}{2\mathcal{H}^2}\frac{\partial^2}{\partial s_\parallel^2}
\left[\left(1+\delta\right)\left(v_\parallel^2+\sigma_{\parallel \parallel}
\right) \right]+...
\label{eqredshiftspace}
\end{equation}
The linearized version of this equation, with zero-order velocity dispersion
$\bar{\sigma}_{\parallel \parallel}=\mathcal{H}^2 \tsig_0$, 
is
\begin{equation}
\delta_s\left(\vs\right)=\delta\left(\vs\right)-\frac{1}{\mathcal{H}}
\frac{\partial v_\parallel}{\partial s_\parallel}
+\frac{\tsig_0}{2}\frac{\partial^2\delta}{\partial s_\parallel^2} +
\frac{1}{2\mathcal{H}^2}\frac{\partial^2\sigma_{\parallel \parallel}}{\partial s_\parallel^2}
\end{equation}
The first two terms are the usual linear theory result
\citep{1987MNRAS.227....1K}, while the last two are related to the inclusion of
dispersion.
In Fourier space, and using $v^i=-i\mathcal{H}k^i k^{-2}\theta$ and
$\sigma_{ij}=2 \mathcal{H}^2 \tsig_0 k_i k_j k^{-2}\pi$ (for irrotational 
velocities), this is
\begin{equation}
\delta_s\left(\vk\right)=
\left(1-\frac{1}{2}k^2\tsig_0~\mu^2\right)\delta\left(\vk\right)+
\mu^2~ \theta\left(\vk\right) - k^2\tsig_0~ \mu^4~ \pi\left(\vk\right)  ~.
\end{equation}
where I have not used the usual $\theta = f\delta =
\frac{d \ln D}{d \ln a}\delta$ simplification because the relation between 
$\theta$ and $\delta$ is complicated by the presence of dispersion (their 
Jeans smoothing kernels are different --
note that, like \cite{2009PhRvD..80d3504P}, I found in Fig. 
\ref{figkerns} that the 
velocity divergence is suppressed 
substantially more than the density fluctuations, which means that these 
corrections will 
be more important for redshift space distortions than they would be in real 
space).
Similarly, $\pi$ will only be 
proportional, not equal, to $\delta$ on large scales (usually smaller), and 
will have a different smoothing kernel. 
The first dispersion term looks like the first term in 
a small-$k$ expansion of the smoothing kernels that have been used to model 
non-linear peculiar velocities in the past 
\cite{2004ApJ...617..782J}. There will clearly be a 
series of similar higher order terms that could be re-summed into a simple
smoothing kernel. Note, however, that there will also be a series of terms 
involving the homogeneous kurtosis of the velocity distribution, which 
generally 
could be of similar order to the re-summed terms, so it isn't obvious that 
the re-summation will be particularly useful (i.e., once the re-summed terms 
are needed, other independent terms may be needed as well).   
Also, note that in cases
where the velocity dispersion is relatively small, the similar distortion 
coming from the
$v_\parallel^2$ term in Eq. (\ref{eqredshiftspace}) could easily dominate
over this term. 

Clearly, redshift-space distortions will require further consideration as the 
higher
order program suggested by this paper is carried out. The suppression of the
new terms by $k^2\tsig_0$ probably means that they will not be significant 
unless terms higher order in $\delta$ are also significant on the same scale, 
i.e., 
$k^2\tsig_0$ probably should be considered to be a new expansion parameter, at
least as small as $\delta$, and probably $\delta^2$ (this is consistent
with the way that $\tsig_0$ is derived, as proportional to the linear power
spectrum, however, the correspondence is not unambiguous because of the 
very different origin and form of growth of the terms). In any case, 
interpretation of very high precision observations, on the imperfectly linear 
scales where most of the information resides, will clearly require a 
more subtle calculation than has so far been done, however, one should not 
despair, or flee to relying entirely on numerical simulations, because an 
understanding of the detailed redshift-space structure of observations will 
ultimately make our measurements of fundamental physics much more 
robust and believable. Note that nothing requires, or even really suggests,
that the effective $\tsig_0$ for galaxies 
\cite{2006MNRAS.366.1455S,2006MNRAS.368...85T,2006MNRAS.368...37L,
2007MNRAS.374..477T,2007ApJ...659..877T} must be 
equal to that for dark 
matter -- in the end, it will probably be a completely free parameter for
galaxies, like bias.

\section{Discussion \label{secdiscuss}}

The deficiencies of standard perturbation theory are a lack of control of 
the higher moments of the velocity distribution function, beyond the 
pressureless fluid
approximation, problems with accuracy related to the fact that PT integrals
include small scales that are generally highly non-linear, even when the 
scales we are interested in are weakly non-linear, and, of course, simply not
working on small scales where the fluctuations are large. 
The deficiencies of numerical simulations are speed, and the related fact that 
statistics (e.g., the power spectrum) cannot be computed directly, but only as
averages over realizations of the random density field, which must contain many
orders of magnitude more degrees of freedom (e.g., a billion) than one really 
needs to describe the statistic of interest (e.g., a few dozen for the power
spectrum).
Additionally, and probably most importantly, the cumbersome, opaque nature of
simulation results is greatly exacerbated when they are used to model galaxies
or other observables
instead of just dark matter, while the advantage of being more or less exact
for gravity alone
(at least in the straightforward limits of large box size and particle 
density) is lost.

This paper directly addresses the PT deficiency of missing higher moments of 
the velocity distribution function, showing how they can be re-activated and
generated at a significant level,
starting from first principles. The approach here may also improve PT by 
providing natural regulation of small-scale structure, i.e., the Jeans 
smoothing that arises here has the same kind of effect as the 
propagator resummation of \cite{2006PhRvD..73f3520C,2006PhRvD..73f3519C}.
The philosophy of this paper is that the cold ``streams'' often discussed as 
``crossing'' are mythical objects -- what one sees in the real Universe is 
always some evolution of effectively warm (although maybe not very warm) 
material. 
The meaning of this idea is very clear for power 
law power spectra, where, as we saw, non-trivial effects of velocity dispersion
can be computed without any initial dispersion entering the discussion. The 
dispersion bootstraps itself up from an arbitrarily small start. The 
temperature is locked into a sort of equilibrium with the growth of structure.
The situation is not as clear for $\Lambda$CDM power spectra, not so much 
because the 
$\Lambda$CDM is cold as because there is very little small-scale power in 
these models,
so the dispersion computed to linear order in the power spectrum remains well
below the non-linear scale (although orders of magnitude larger than it would
be if there was no structure). This situation will change when calculations are
done to higher order, where I showed that there is enough small-scale power 
generated to 
produce dispersion that would be far too large in absence of feedback on the 
structure formation itself.
The initial velocity dispersion should be rendered irrelevant when the system 
moves 
into a sort of self-regulating mode, like the power law example, where the 
velocity dispersion and growth
of structure are tightly coupled by feedback between them. Ultimately,
it seems likely that the best effective small-scale model for doing 
perturbation theory will involve some more general balance between different 
pieces,
i.e., density, velocity divergence, dispersion, and maybe even vorticity, etc., 
because we
know that physically this is what the small-scale structure really is, i.e., 
halos which 
can only be described as a delicate balance of the original 
perturbative LSS quantities.
To put this another way: hopefully, when all of the relevant elements are 
included, there will  be some fixed-point structure for small scales with a 
clear physical interpretation and
effectively far fewer than the original number of degrees of freedom (akin to
the fixed-point power law found in \cite{2007PhRvD..75d3514M}, but with richer 
structure).

This paper does not exactly contain useful quantitative take-home results. 
The results are 
primarily a procedure to follow for future calculations. 
If there is a single equation that 
best represents the results, it is probably Eq. (\ref{eqkNLsigRG}), which
shows how, for a power law power spectrum, the velocity dispersion tracks the
non-linear scale, with Jeans filtering erasing more and more small-scale 
structure as the larger scale structure grows. 
Eqs. (\ref{eqTnonlinear})
and (\ref{eqtsigevolution}) are also critical,
in that they show generally how homogeneous velocity dispersion is generated 
out of fluctuations. The key new variable, equivalent 
to $\delta$ and $\theta$, but for perturbations in velocity dispersion, is 
$\pi \propto \partial_i \partial_j \sigma^{ij}$.

The next step in this line of work is to derive to the 
next-order equations 
like Eq. (\ref{eqPddevolution}-\ref{eqPppevolution}) (but including bispectrum
equations), which are needed to have
any chance of properly describing $\Lambda$CDM. Then the results can be tested
by comparison to numerical simulations. While $\Lambda$CDM is of course the 
ultimate goal, tests of the theoretical concepts here might be more 
conveniently done with 
power law simulations, particularly ones with relatively blue spectra, as in
\cite{2009MNRAS.397.1275W}. 
Any of the renormalization approaches that have been developed recently can
probably be applied to the problem of renormalizing the velocity dispersion,
at least in principle, i.e., resumming a set of terms that generates the 
dispersion, either explicitly or through a renormalization group equation.
The obstacle to doing this elegantly may be the difficulty of 
obtaining analytic solutions as
the evolution equations become more complicated (this problem of requiring 
analytic solutions is, I think, the 
primary reason to favor the ``numerical evaluation of evolution equations for 
statistics'' approach 
advocated in this paper over other, more completely analytic, recent 
approaches).

One might ask ``why stop with the 2nd moment of the distribution function, 
i.e., why can we drop $q_{ijk}$ from Eq. (\ref{eqdispersion})?''
One hope, which will need to be verified by future
calculations, is that the effect of increasingly high moments on large 
scales may take the form of a gradient expansion, i.e., in Fourier space a 
series where the effects of increasingly high moments enter multiplied by 
increasing powers of $k$.  In this case, we would expect that, on scales where
the effect of the 2nd moment are already small, the effects of higher moments
will be even smaller.          

The result for redshift-space space distortions (Eq. \ref{eqredshiftspace}) 
leads to the question:  Why do we use $\vv$, and in this paper 
$\sigma_{i j}$, as the variables to be solved for perturbatively, rather than,
e.g., momentum 
$\left(1+\delta\right)\vv$ and kinetic energy
$\left(1+\delta\right)\left(v_i v_j+\sigma_{ij}\right)$? An answer one might 
have considered was that velocities are needed to compute redshift-space 
distortions, but 
here we see that the most direct quantities for that are in fact momentum and 
energy, with additional perturbative calculations needed when starting from 
$\delta$, $\vv$, and $\sigma_{ij}$. 
A change to total kinetic energy would have the potentially substantial effect 
of increasing the 
Jeans-like smoothing of the linear power, because the zero order pressure would
be larger.  I do not see any clear a priori reason to favor one option over 
another 
-- they are just different ways of arranging a series of terms, which should be
equivalent if one could include an infinite number of terms.  Note that, while
the choice of variables is optional, the renormalization of the energy-related
variable is not optional -- it simply makes no sense to ignore the fact that 
the size of terms in the series describing one of your variables is 
increasing rapidly, rather than decreasing, with order. 
There is a lot of circumstantial evidence that using total kinetic energy could
be useful.  The renormalization/resummation of the propagator in 
\citep{2006PhRvD..73f3520C} leads to
filtering much like the Jeans filtering we find due to velocity dispersion, 
but with scale given by the bulk velocity power spectrum.  The Lagrangian
PT-based scheme of \citep{2008PhRvD..77f3530M} leads to a similar result.
Another possibility to consider would be the evolution of large-scale fields  
with the 
small-scale structure explicitly averaged out, which would naturally lead to 
the inclusion of small-scale bulk velocities in the dispersion term
\cite{2000PhRvD..62j3501D,2002MNRAS.334..435D,2003AN....324..560D}.
One might also consider using the Schr\"{o}dinger equation representation of
the exact Vlasov equation, proposed in \cite{1993ApJ...416L..71W}, combined
with the approach of this paper.

In the end, one could view the approach in this paper as a re-activation 
of the program under discussion in papers like 
\cite{1977ApJS...34..425D}, which set out to integrate the BBGKY equations
numerically.  This reconsideration is timely because of several developments
over the last thirty years.  We now know the appropriate class of models
to focus on, especially including the appropriate the initial conditions.  
There has been a lot of work on both perturbation theory beyond linear order,
and on N-body simulations, with the limitations of each teaching us a lot about
what is needed from new methods.
We also have a specific calculation to focus on:  the scales where baryonic
acoustic oscillations (BAO) are 
observable \citep{1998ApJ...496..605E,
2003ApJ...594..665B,2003PhRvD..68h3504L,2003ApJ...598..720S,
2004ApJ...615..573M,2005ApJ...631....1G,
2005MNRAS.357..429A,2005MNRAS.363.1329B,2006MNRAS.365..255B,
2007PhRvD..76f3009M}, 
which points us toward the perturbative approach that motivates truncating the
hierarchy and believing that high precision can be achieved (in contrast to 
\cite{1977ApJS...34..425D}, who were focused on the more strongly non-linear
regime, where the truncation used here is not well-motivated). 
The same scales are also potentially the most powerful for other
measurements based on, e.g., redshift-space distortions 
\cite{2008arXiv0810.0323M}. 

The basic form of calculation I do here is completely standard in some other 
areas of cosmology.  For example, the evolution of the homogeneous (background)
value of an interacting scalar field in the early Universe is affected by 
quantum and thermal fluctuations.  Its evolution is not described by the 
original equation of motion with all perturbations ignored, but by a 
renormalized effective potential, which is at least 
formally infinitely different from the original potential, i.e., completely
dominated by the part due to fluctuations \cite{2005pfc..book.....M}.
Another interestingly similar calculation is the development of equilibrium 
after preheating after inflation. \cite{2000hep.ph...11159F,
2001PhRvD..63j3503F,
2006JCAP...07..006D,2006PhRvD..73b3501P,2007PhRvD..75d3518F} perform fully
non-linear simulations of the interaction of scalar fields, much like 
large-scale structure simulations, with the
added twist that the background density and pressure are affected by the 
fluctuations.  The evolution of the scale factor is calculated by taking
spatial averages over the fluctuations as the simulation is running.  In the
beginning, the nearly homogeneous inflaton field dominates, and a very naive
calculation might compute the expansion of the simulation in advance assuming 
homogeneous evolution, but by the end of the simulation the homogeneous 
component has actually practically disappeared, with the background evolution 
completely dominated by the fluctuations, which behave like radiation.  
Of course no one would ever do the very naive
calculation just mentioned, where the affect of the fluctuations on the 
background equation of state is ignored... except that this is what we do when 
we assume 
that the tiny initial temperature of CDM means that it will forever remain 
cold.

Substantial work remains to determine if the approach in this paper enhances 
the
accuracy of predictions of quasi-linear clustering in the real Universe;
however, it is now possible to consider an effect that was previously outside
of any first-principles computational control in this kind of perturbation 
theory.
The bottom line of this paper is that even linear theory fluctuations 
necessarily imply a significant
one-loop renormalization of the background velocity dispersion.

\acknowledgements

I thank Roman Scoccimarro for helpful comments on the manuscript and Lev Kofman 
for helpful conversations.
I acknowledge the support of the Beatrice D. Tremaine Fellowship.

\bibliography{$LATEX/cosmo,$LATEX/cosmo_preprints}

\end{document}